\documentclass[prc,aps,superscriptaddress,nofootinbib,twocolumn,showkeys]{revtex4}
\usepackage{amssymb}
\usepackage[dvips]{graphicx}
\usepackage[english]{babel}
\usepackage{indentfirst}
\usepackage{amsxtra}
\usepackage{amsmath}
\usepackage{supertabular}
\usepackage{multirow}
\usepackage[mathcal]{eucal}
\usepackage{cancel,soul,ulem}
\usepackage[usenames]{color}
\usepackage{ulem}

\begin{document}
\title {\textbf{Towards a power counting in nuclear energy--density--functional theories through a perturbative analysis}}
\author{Stefano Burrello}
\affiliation{Universit\'e Paris-Saclay, CNRS/IN2P3, IJCLab, 91405 Orsay, France}
\author{Marcella Grasso}
\affiliation{Universit\'e Paris-Saclay, CNRS/IN2P3, IJCLab, 91405 Orsay, France}
\author{Chieh Jen Yang}
\affiliation{Department of Physics, Chalmers University of Technology, SE-412 96, G{\"o}teborg, Sweden}

\begin{abstract}
We illustrate a step towards the construction of a power counting in energy--density--functional (EDF) theories, by analyzing the equations of state (EOSs) of both symmetric and neutron matter. Within the adopted strategy, next--to--leading order (NLO) EOSs are introduced which contain renormalized first--order--type terms and an explicit second--order finite part. Employing as a guide the asymptotic behavior of the introduced renormalized parameters, we focus our analysis on
two aspects: 
(i)
With a minimum number of counterterms introduced at NLO, we show that each energy contribution entering in the EOS has a regular evolution with respect to the momentum cutoff (introduced in the adopted regularization procedure) and is found to converge to a cutoff--independent curve. The convergence features of each term are related to its Fermi--momentum dependence. 
(ii) 
We find that the asymptotic evolution of the second--order finite--part coefficients is a strong indication of a perturbative 
behavior, which in turns confirms that the adopted strategy is coherent with a possible underlying 
power counting in the chosen Skyrme--inspired EDF framework.
\end{abstract}

\maketitle

EDF theories, developed over the last decades for the nuclear many--body problem, have been built in most cases on the basis of a phenomenological methodology. 
Differently than density--functional theories (DFTs) \cite{HK,KS,kohn,dreizler,parr,takao}, 
employed in chemistry and in solid state physics, they were not generated starting from founding theorems, such as the Hohenberg-Kohn theorems in DFT \cite{HK,kohn}. In addition, they were not formulated within a well--defined general theoretical scheme (for instance, within a low--density Fermi gas expansion \cite{LY,yang,bishop,gali,abri,fw}  or within the Dyson many--body perturbative expansion \cite{fw})  in such a way that, order by order, one could expect systematic and controlled improvements in the theoretical models. The need for a more fundamental methodology gave a push in the last years towards the design of functionals inspired by effective--field theories (EFTs) and/or benchmarked on {\it{ab--initio}} models \cite{grasso2019,yglo,elyo1,drop,elyo2,papa1,boulet,papa2,papa3,lacroix1,lacroix2,roggero,holt,navarro,holt1,kaiser,bogner,drut}.

In particular, contrary to what happens in EFTs, a power--counting scheme, providing a hierarchy  of the interaction contributions, is still missing in EDF theories, due to the lack of an {\it{a priori}} identification of a breakdown scale. 
The development of functionals based on a power counting would represent a crucial achievement in EDF, for example within the perspective of using such functionals in sophisticated beyond--mean--field (BMF) models, where the leading order (LO) of the Dyson equation is overcome \cite{grasso2019}. 

EDF theories, based for example on Skyrme interactions \cite{sk1,sk2,VB}, are globally very successful in describing ground-- and excited--states phenomenology, as well as reactions, over the entire chart of nuclei. This success should imply that there must be an associated EFT with some defined power counting. However, one does not know {\it{a priori}} anything about such an EFT and the related power counting. To make some progress, one can assume a specific scheme  based on a chosen starting point (for example guided by renormalizability and by the known properties of Skyrme interactions) and check the obtained results to assess whether the assumed scheme is coherent with a possible underlying power counting. This is the strategy that we follow.

At densities of interest for finite nuclei, which define the range of validity of the scheme that we want to construct, we cannot employ an EFT treatment valid for dilute Fermi gases, for which an expansion parameter is known 
\cite{HF}. Guided by the fact that mean--field (MF) theories are successful in EDF and that, on the other side, BMF theories are able to provide even better results than MF for many nuclear phenomena, we employ an order--by--order development which is dictated by the many--body Dyson expansion. 
Such a development has specific leading and subleading Feynman diagrams: LO corresponds to the MF approximation, NLO to the second--order approximation, and so on. 
We do not know what is our EFT expansion parameter, which is to be extracted afterwards (for example after a full power--counting analysis is done with a Lepage--like plot \cite{Lepage,harald,harald2}). Unfortunately, due to the fact that there is no cutoff dependence at LO, we cannot assess its residual cutoff dependence. As a consequence, the breakdown scale, which can be indicated by the crossing point in a Lepage plot from two--different--order curves, cannot be extracted at this stage. This will be performed once next--to--next--to--leading order (NNLO) contributions will be also evaluated, by comparing NLO and NNLO results.

But this does not prevent us from starting an EFT procedure, based on the Dyson expansion. Guided by the requirement of renormalizability and choosing to work with a Skyrme--type interaction, we may identify more precisely our LO interaction, which is then chosen as the so--called $t_0-t_3$ model.  This choice is also justified by the fact that such a model represents the minimum form of the effective interaction necessary to produce an equilibrium point in symmetric matter at the MF level. One can thus imagine that many essential properties are already captured. The traditional approach for Skyrme effective interactions is then to include more terms such as gradient terms, so to improve the description of both matter and nuclei. However, from conventional EFT (such as chiral EFT), it is possible that those corrections should enter as subleading contributions. Thus, how do we improve EDF in a controlled manner? 

In Ref.  \cite{yang2017-b} we evaluated the second--order contribution based on the chosen LO effective interaction ($t_0-t_3$ model) and were insured that results satisfy up to NLO the renormalizability condition \cite{yang2017-a}. 
In the present work, with a more careful systematic optimization protocol of the low--energy constants, we have found a strong indication of a perturbative behavior: after renormalization, the finite part obtained from second--order diagrams (which can be regarded as the true physical correction to LO) contributes much less than the LO to the EOS. This is a crucial result, since it is a confirmation that we are moving towards a reasonable direction with the choice of such an LO interaction. If we had found that the NLO finite--part contribution was greater than the LO contribution, we should have simply concluded that this choice had to be excluded, the whole scheme to establish EFT from EDF with the chosen LO effective interaction being not converging.

The perturbative behavior will be shown to be even more pronounced in the low--density regime of matter, opening interesting perspectives for future applications to exotic nuclei with dilute surfaces and to neutron--star crusts \cite{burPRC92, burPRC94}. The regime of validity of the proposed scheme should be intended to be the one typically covered by most EDF theories, that is from densities below the saturation of symmetric matter up to around two or three times the saturation density itself, where internal degrees of freedom of nucleons may still be safely disregarded. At the same time, as for many existing EDF models, the description of extremely low densities is not the scope of the proposed scheme. 

We do not claim to provide here the final solution to this open problem of defining a power counting within EDF theories, since the present study does not yet permit us to carry out uncertainty analyses and to associate errors to the chosen expansion truncation. Nevertheless, from the found perturbative behavior and under the assumption that the underlying breakdown scale is at least the Fermi momentum at twice the saturation density (the assumed range of validity of our theory), we can conclude that the proposed scheme is coherent with a possible underlying power counting.

An interaction $V_{LO}$ is introduced, 
given by the $t_0-t_3$ part of a zero--range Skyrme interaction,  
\begin{eqnarray}
V_{LO}&=& t_0(1+x_0P_{\sigma})
+\frac{t_3}{6} (1+x_3P_{\sigma}) \rho^{\alpha},  
\label{vlo}
\end{eqnarray}
where $t_0$, $t_3$, $x_0$, $x_3$, and $\alpha$ are parameters, $\rho$ is the density, and $P_{\sigma}=(1+{\sigma}_1 {\sigma}_2)/2$ is the spin--exchange operator.  
Using natural units $\hbar=c=1$, the associated LO EOSs are the mean--field (first--order) EOSs for symmetric matter (SM) and neutron matter (NM),  
\begin{equation}
\frac{E_{SM}^{\rm (LO)}}{A}=\frac{3k_{F}^{2}}{10m}+\frac{t_0}{4 \pi ^{2}}k_{F}^{3}+\frac{t_{3,\alpha}}{4\pi^2}
k_{F}^{3+3\alpha}  \label{meanfieldsm}
\end{equation}%
and 
\begin{eqnarray}
\frac{E_{NM}^{\rm (LO)}}{N} &=&\frac{3k_{F}^{2}}{10 m}+\frac{t_0^*}{12\pi ^{2}%
} k_{F}^{3}  
+\frac{t_{3,\alpha}^*}{12 \pi^2} k_{F}^{3+ 3\alpha },  \label{meanfieldnm}
\end{eqnarray} 
respectively, where $t_{3,\alpha}=(2/(3\pi^2))^{\alpha} t_3/6$, 
$t_0^*=t_0(1-x_0)$,  
$t_{3,\alpha}^*=(1/(3\pi^2))^{\alpha} t_3^*/6$, 
and $t_3^*=t_{3}(1-x_{3})$. 
The Fermi momentum $k_F$ is related to the density by the relation 
$k_{F}=(3\pi^{2}\rho/2)^{1/3}$ [$k_{F}=\left( 3\pi ^{2}\rho\right) ^{1/3}$] for SM [NM]. 

The NLO EOSs consist of the contributions obtained with the once--iterated LO interaction $V_{LO}$ \cite{mog,yang2016} plus selected counterterms, 
used to absorb some of the ultraviolet divergences induced at second order by the zero--range LO interaction. 
The counterterms are associated to the so-called $V_{NLO}$ interaction, which 
is treated as a first--order contribution in the NLO EOS (Fig. 3 of Ref. \cite{yang2017-b}).
Note that three divergences --- each with a different power of $k_F$ --- occur at second order, only two of them having  $k_F$--dependencies already present in the MF EOS. Therefore, to ensure renormalizability, it is necessary to adopt at least one counterterm at NLO to eliminate the divergence associated with the new $k_F$ dependence. In Ref. \cite{yang2017-b}, the introduced parameters were adjusted  to  reproduce the benchmark EOS within three scenarios, which differ on the number of counterterms. However, without requiring any constraints on the parameters (and given their large number), important fluctuations were found in their evolution as a function of the momentum cutoff $\Lambda$ introduced in the regularization procedure. 
Guided by the asymptotic behavior of the introduced renormalized parameters and performing a finer adjustment procedure, we find here a quite regular evolution of the interaction parameters as a function of the cutoff. This allows us to study their cutoff dependence as well as that of the different contributions to the total energy.

We work within what was called scenario (c) in Ref. \cite{yang2017-b}, which implies a minimal number of couterterms (one type), introduced to handle the second--order terms having a linear divergence with respect to $\Lambda$ and a $k_F$ dependence of the type $k_F^{3+6\alpha}$. 
Such divergent terms are regrouped with the counterterm  and this contribution enters in the EOS with a renormalized parameter $C^{\Lambda}$ ($C^{\Lambda*}$) for SM (NM). The other second--order linear divergences, of the type $k_F^{3}$ and $k_F^{3+3\alpha}$, are  absorbed in MF terms (Eqs. (\ref{meanfieldsm}) and (\ref{meanfieldnm})), generating renormalized parameters $t_{0}^{\Lambda}$, $t_{0}^{\Lambda*}$, $t_{3,\alpha}^{\Lambda}$, and $t_{3,\alpha}^{\Lambda*}$. 
We limit ourselves to the scenario (c) also because, assuming an underlying EFT--type expansion truncated at the order $(\frac{Q}{\Lambda_{hi}})^2$  ($Q$ and $\Lambda_{hi}$ denoting, respectively, the nucleon momentum and the breakdown scale of the expansion \cite{yang2017-b}), 
 the comparison among the three scenarios have indicated in Ref. \cite{yang2017-b}
that their difference is likely to be non negligible only at higher order.
The NLO EOSs for SM and NM are, respectively, 
\begin{widetext}
\begin{eqnarray}
\frac{E_{SM}^{\rm (NLO)}(k_{F},\Lambda)}{A} =\frac{3k_{F}^{2}}{10m}+\frac{%
t_{0}^{\Lambda}}{4\pi ^{2}} k_{F}^{3}+\frac{t_{3,\alpha}^{\Lambda}}{4\pi ^{2}} k_{F}^{3+3\alpha} +%
\frac{C^{\Lambda}}{4\pi ^{2}}k_{F}^{3+6\alpha }  + 
F_{S,1}^{(2)} k_{F}^4 + F_{S,2}^{(2)} k_{F}^{4+3\alpha} + F_{S,3}^{(2)} k_{F}^{4+6\alpha}  
  \label{a}
\end{eqnarray}
and 
\begin{eqnarray}
\frac{E_{NM}^{\rm (NLO)}(k_{F},\Lambda)}{A} =\frac{3k_{F}^{2}}{10m}+%
\frac{t_{0}^{\Lambda*}}{12\pi ^{2}} k_{F}^{3} +\frac{t_{3,\alpha}^{\Lambda*}}{%
12 \pi^2} k_{F}^{3+3\alpha}  
+\frac{C^{\Lambda\ast }}{%
12\pi ^{2}}k_{F}^{3+6\alpha }+
F_{N,1}^{(2)} k_{F}^4 + F_{N,2}^{(2)} k_{F}^{4+3\alpha} + F_{N,3}^{(2)} k_{F}^{4+6\alpha},
  \label{a2}
\end{eqnarray}%
\end{widetext}
where the cutoff $\Lambda$ is put on the outgoing relative momentum $\vec{k}'=(\vec{k}_1'-\vec{k}_2')/2$ ($\vec{k}_{1(2)}'$ being the outgoing momentum of the nucleons 1 (2)). 
The only explicitly appearing second--order terms are the last three terms of Eqs. (\ref{a}) and (\ref{a2}), which are  the 
second--order finite parts. 
The expressions for the renormalized parameters and for the quantities $F^{(2)}$ are provided in Appendix \ref{notation}.
One may notice that the second--order finite parts and the counterterm contributions have higher powers of $k_F$, 
compared to the renormalized ($t_0^{\Lambda},t_{3,\alpha}^{\Lambda}$) model, if $0 < \alpha < 1/3 $, which will be always the case in our adjustments. 
The rearrangement terms are included as in Ref. \cite{yang2017-b} (see Refs. \cite{pastore,carl}). 
In this work, the value $m=939$ MeV is used and the benchmark curves on which the adjustments are carried out are the SM and NM SLy5 \cite{cha1,cha2,cha3} mean--field EOSs. These are reasonable EOSs for both SM (reproducing the empirical saturation point) and NM (adjusted on a microscopic EOS). We stress that we could have chosen other benchmark EOSs. Any other choice of reasonable EOSs for both SM and NM would not change the generality of our discussion. 

The adjustment of the parameters $t_0$, $t_3$, $x_0$, $x_3$, $\alpha$, $C^{\Lambda}$,  and 
$C^{\Lambda*}$
is done on $N$ = 11 points of the SLy5 SM and NM EOSs, using cutoff values ranging from  2 to 30 fm$^{-1}$. The $\chi^2$ values to be minimized are computed as $\chi^2=1/(N-1)\sum_i^N (E_i -E_{i,ref})^2/\Delta E_i^2$, where  $E_{i,ref}$ are the reference values on the chosen benchmark curves, and $\Delta E_i$ are taken equal to 1\% of the reference values. 
We choose not to perform at this stage a detailed uncertainty quantification such as a Bayesian analysis \cite{bay,bayb,bayc,bayd,baye,bayf,bayg}, because our benchmark is chosen between (many) possible existing reasonable EOSs for matter. A more systematic uncertainty quantification will be carried out when also finite nuclei will be treated.
  
Nevertheless, we report in Table \ref{fitc2} in Appendix \ref{adjust}
the $\chi^2$ values together with 
the adjusted parameters. 
We checked that their values are stable when varying from 9 to 13 the number of points and that some spurious correlations intrinsically existing among some parameters (for example between $x_3$ and $\alpha$) do not qualitatively affect our conclusions. 
In particular, one observes that $\alpha$ increases monotonically and tends to saturate to a value smaller than 1/3. 
The quality of the fit is instead slightly deteriorated when constraining $\alpha$ to a fixed value, which lies within its explored range of variation. 

This aspect should deserve a deeper analysis. In EDF theories we are used to work with an adjustable parameter $\alpha$ in order to absorb in an effective way several effects, for instance $A$-body effects. However, from an EFT perspective, if an underlying expansion exists, $\alpha$ should not probably play the same role and should be a multiple of 1/3, coherently with a $k_F$ expansion. In this sense, we are not fully satisfied by the pragmatic choice that we adopt at this stage, that is keeping $\alpha$ as a free parameter. Maybe, future higher-order calculations at NNLO will give us additional hints for addressing this point more properly by using for $\alpha$ a fixed value. 

In light of the performed adjustement procedure, 
we focus our analysis on two main aspects.  

i)  
Each renormalized parameter (each contribution to the total energy) is found to have a convergent behavior when the cutoff increases. 

ii)
We address the issue of  perturbativity, which is crucial in our framework to have an order--by--order convergent pattern.

\begin{figure}
\includegraphics[scale=0.34]{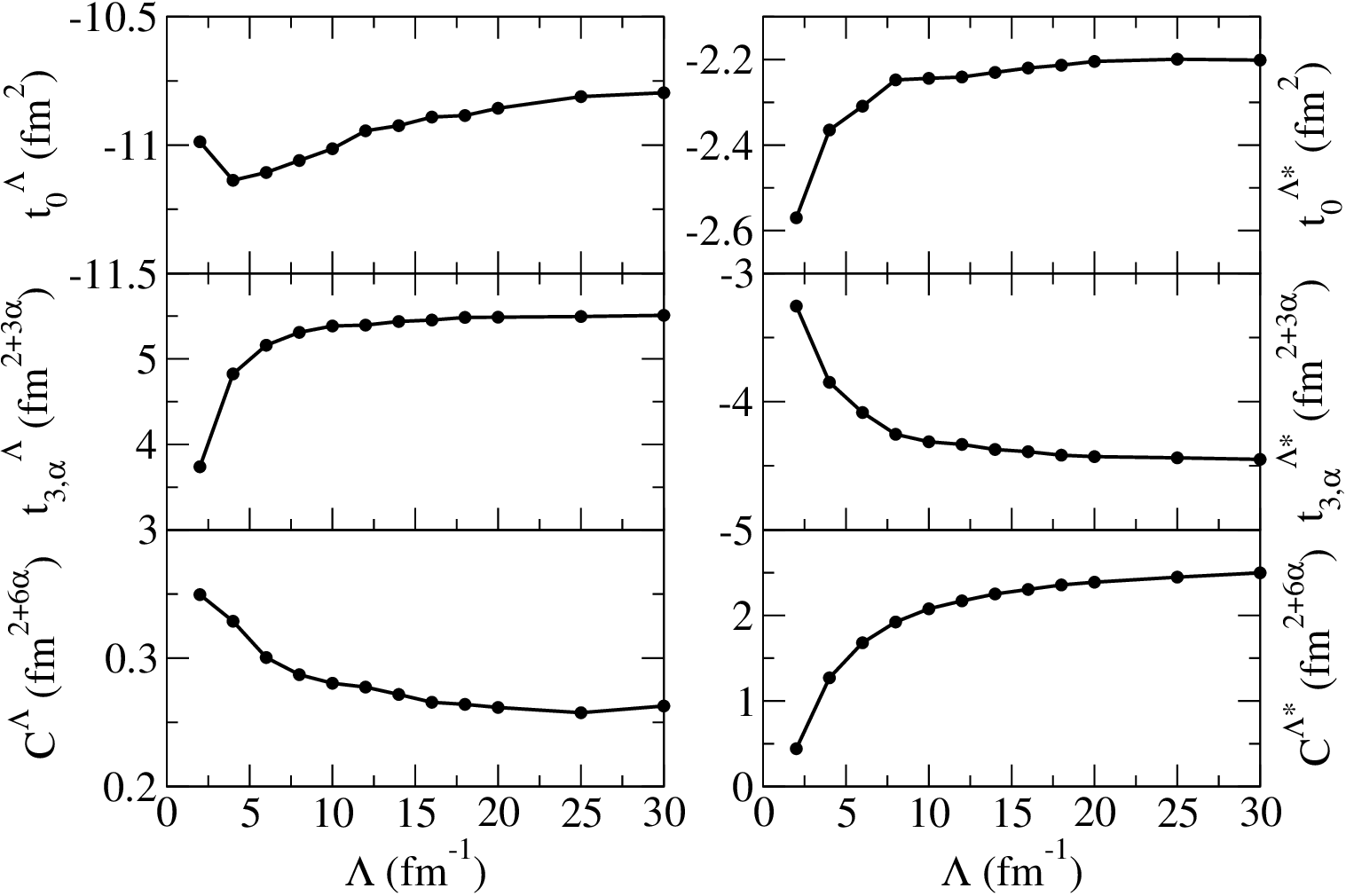}
\caption{Renormalized parameters as a function of the cutoff.}
\label{paramet}
\end{figure}

Figure \ref{paramet} shows the trend of the renormalized parameters as a function of $\Lambda$. They show a regular evolution and a convergent behavior as a function of the cutoff. A similar pattern was observed in the renormalization of pionless \cite{bira9808007,bira99,pionless,pionlessa,pionlessb,pionlessc,pionlessd,pionlesse,pionless1} and chiral EFT \cite{nogga,Ya09A,Ya09B,ZE12}. From Eqs. (\ref{t0})-(\ref{t3tilde}), such a convergent behavior means that, when $\Lambda \rightarrow \infty$, $\Lambda G$, $\Lambda H$, 
$\Lambda I$, and $\Lambda L$ must tend to a finite value. This implies that $G$, $H$, $I$, and $L$ must behave asymptotically as $\sim 1/\Lambda$ and that, when $\Lambda \rightarrow \infty$, $t_0$ and $t_3$ must go to zero as $1/\sqrt{\Lambda}$. 
Such a behavior of $t_3$  in turn constrains 
 $P$ and $R$ (Eqs. (\ref{pp}) and (\ref{qq}) in Appendix \ref{notation}), which must behave asymptotically as $\sim 1/\Lambda$:   $\Lambda P$ and $\Lambda R$ go to a finite value when $\Lambda \rightarrow \infty$. 
Finally, up to an uncertainty of the order $(Q/\Lambda_{hi})^2$, all the coefficients of the finite parts in Eqs. (\ref{ff1}) and (\ref{fff}) are expected to be progressively suppressed as $1/\Lambda$ when the cutoff increases. 

The convergent behavior of the renormalized parameters is indicated by their small variation at large $\Lambda$: indeed, their values at $\Lambda=20$ and 30 fm$^{-1}$
differ by only 0.6, 0.4, 0.4, 0.1, 0.5, and 4.4 \% for $t_0^{\Lambda}$, $t_{3,\alpha}^{\Lambda}$, $C^{\Lambda}$, 
$t_0^{\Lambda*}$, $t_{3,\alpha}^{\Lambda*}$, and $C^{\Lambda*}$, respectively. 
The convergence features are however not the same for all the parameters. 
For SM (NM), variations smaller than 2 (6) \% between two values of renormalized parameters  may be found comparing the values at $\Lambda=$ 4 and 10  fm$^{-1}$  for $t_0^{\Lambda}$ ($t_0^{\Lambda*}$). 
For $t_{3,\alpha}^{\Lambda}$ ($t_{3,\alpha}^{\Lambda*}$), this occurs between 10 and 20 fm$^{-1}$. For $C^{\Lambda}$ ($C^{\Lambda*}$), one needs to go to higher--cutoff regions  
to reach such a flat behavior. 
This is a consequence of the relative importance of the energy contributions at different values of $\rho$. 
At the lowest densities, the most important interaction contribution to the EOS is $E_0$ (depending on $k_F^3$). The adjustment to obtain a $\Lambda$--independent EOS at all cutoff values (even the lowest ones) 
in the lowest--density regions  affects mostly $t_0^{\Lambda}$ and $t_0^{\Lambda*}$ which are then constrained to get closer to a flat behavior starting from small values of the cutoff (to guarantee that the adjusted curve at the lowest densities has the desired behavior at all cutoff values). By increasing the density, the relative importance in the EOS of the terms with increasing powers of $k_F$  becomes progressively higher. This means that, to have a $\Lambda$--independent EOS, the corresponding 
parameters get progressively closer to their last values calculated here (corresponding to $\Lambda=$ 30 fm$^{-1}$) at increasing cutoff values.
 
As one may see in Fig. \ref{paramet}, the renormalized parameters related to the expansion seem to be quite comparable in the entire window of cutoff values shown in the figure. There is a factor of approximately 2 between $t_0^{\Lambda}$ and $t_{3,\alpha}^{\Lambda}$ and between  $t_0^{\Lambda*}$ and $t_{3,\alpha}^{\Lambda*}$. This is an indication of the naturalness of the coefficients,  
 which should be preserved also for cutoff values near the (still unknown) breakdown scale owing to the found convergent behavior of the renormalized parameters with respect to the cutoff. 

\begin{figure}
\includegraphics[scale=0.34]{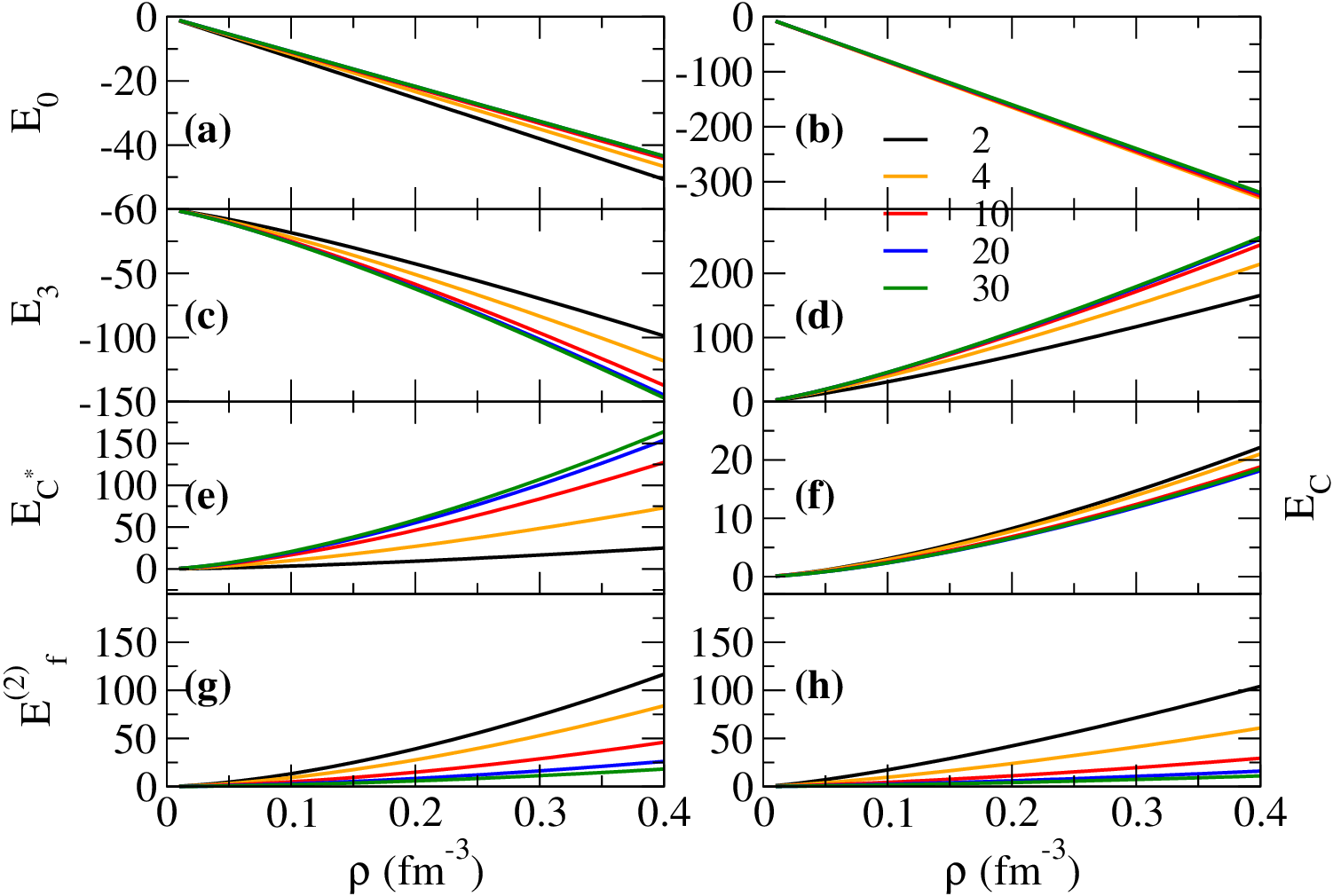}
\caption{(Color online) Energy contributions (in MeV) to the NLO EOSs for NM (panels (a), (c), (e), and (g)) and SM (panels (b), (d), (f), and (h)). The values of $\Lambda$ (in fm$^{-1}$) are shown in the legend. $E_0$, $E_3$, and $E_{C(C^*)}$  
 represent, respectively, the second, the third, and the fourth terms of Eqs. (\ref{a}) and (\ref{a2}). 
$E^{(2)}_{f}$ in panels (g) and (h) is the sum of the  
the three contributions $F_1^{(2)}$, $F_2^{(2)}$, and $F_3^{(2)}$ of Eqs. (\ref{a}) and (\ref{a2}).  }
\label{nmenergy}
\end{figure}

The convergent behavior may be seen in the energy contributions to the EOS, shown in Fig. 
 \ref{nmenergy} 
 for SM (right panels) and NM (left panels), where these energy contributions are plotted as a function of the density for some selected values of the cutoff (see caption). The second--order finite part $E^{(2)}_{f}$ in panels (g) and (h) is the sum of the  
the three contributions $F_1^{(2)}$, $F_2^{(2)}$, and $F_3^{(2)}$ of Eqs. (\ref{a}) and (\ref{a2}).  
 
\begin{figure}
\includegraphics[scale=0.34]{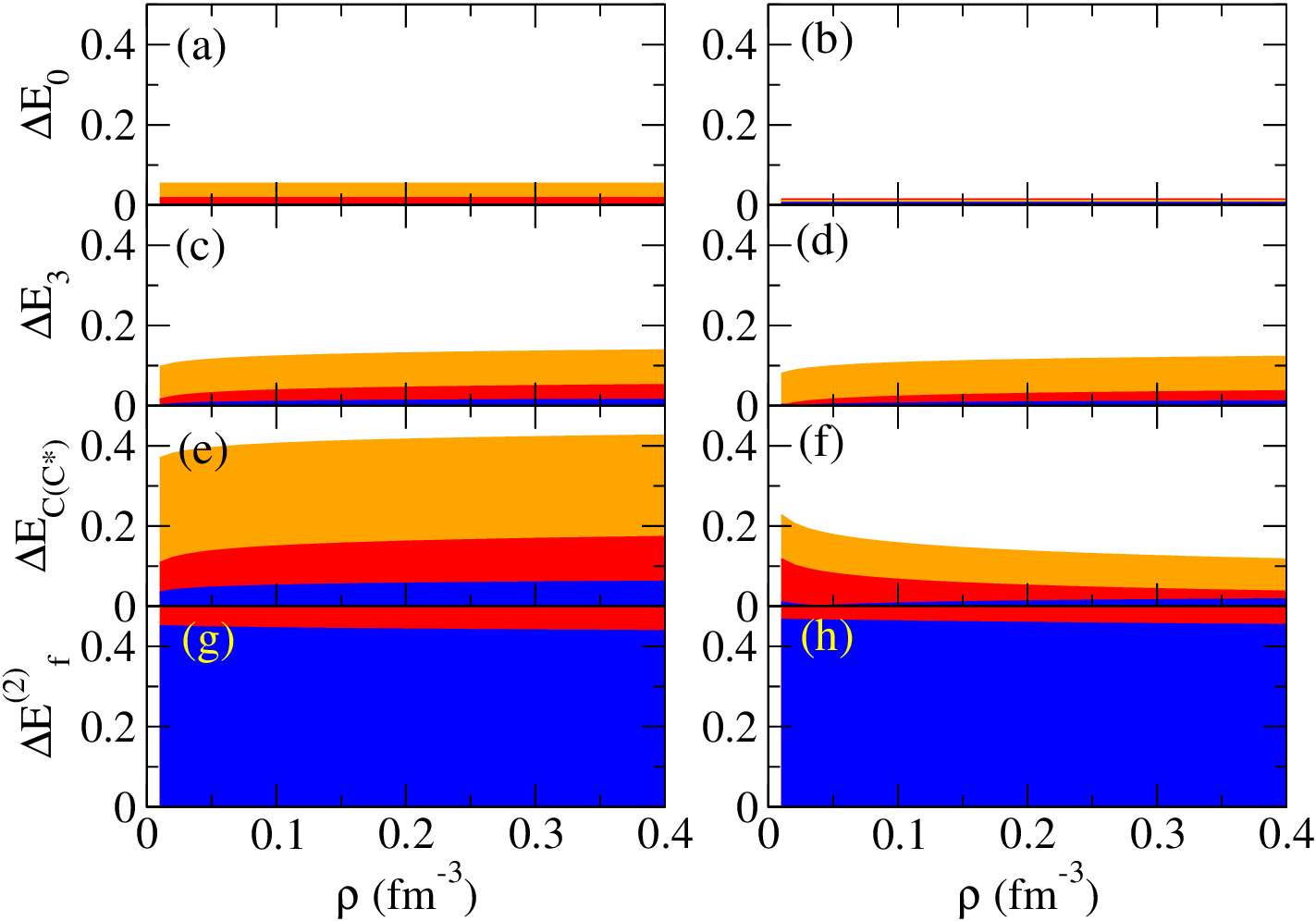}
\caption{(Color online) Orange areas: 
absolute values of the difference between the  energy contributions computed at $\Lambda=$ 4 and 10 fm$^{-1}$, divided by the energy contribution calculated at $\Lambda=$ 10 fm$^{-1}$.   
Red areas: 
absolute values of the difference between the  energy contributions computed at $\Lambda=$ 10 and 20 fm$^{-1}$, divided by the energy contribution calculated at $\Lambda=$ 20 fm$^{-1}$.
Blue areas:
absolute values of the difference between the energy contributions computed at $\Lambda=$ 20 and 30 fm$^{-1}$, divided by the energy contribution calculated at $\Lambda=$ 30 fm$^{-1}$.      
Panels (a), (c), (e), and (g) refer to NM whereas panels (b), (d), (f), and (h) refer to SM.
}
\label{conve}
\end{figure}

Figure \ref{conve} shows the relative variations of the energy contributions (see caption) for both NM (left panels) and SM (right panels). 
One indeed observes similar relative variations compared to those already discussed for the corresponding renormalized parameters.  We may also notice that 
the relative variation of the finite part is still quite large between 20 and 30 fm$^{-1}$. 
In all panels (except panels (a) and (b)) we observe a density dependence. This is due to the presence of $\alpha$ 
in the powers of $k_F$ of these contributions to the EOS: $\alpha$ increases monotonically as a function of the cutoff. 
The range of variation of $\alpha$ can be ascribed to the fact that, when the cutoff increases, the counterterm contribution to the NM EOS must compensate, especially at the highest densities, the fact that the repulsive finite part (containing in principle the highest power of $k_F$) is progressively suppressed. Therefore, when the cutoff is raised, the counterterm contribution has a $k_F$--power in the EOS which correspondingly increases between 4 and 5. 
We remind that such a repulsive contribution is provided, in a MF NM EOS computed with a full Skyrme interaction, by the so--called gradient terms which produce a $k_F^5$--contribution in the EOS. Similar effects generated by  the gradient terms in a traditional MF calculation are then associated in our framework to the counterterm contribution. This also justifies why the Skyrme gradient terms ($t_1$ and $t_2$) are not necessary in our scheme for matter: their inclusion would lead to overconstrained parameters and we checked that the fit quality is not improved when they are taken into account.  In future applications to finite nuclei up to NLO it will be possible to assess whether the functional terms related to the counterterms are able also in that case to reproduce the effects usually generated by gradient terms or if gradient terms have to be explicitly introduced.

\begin{figure}
\includegraphics[scale=0.34]{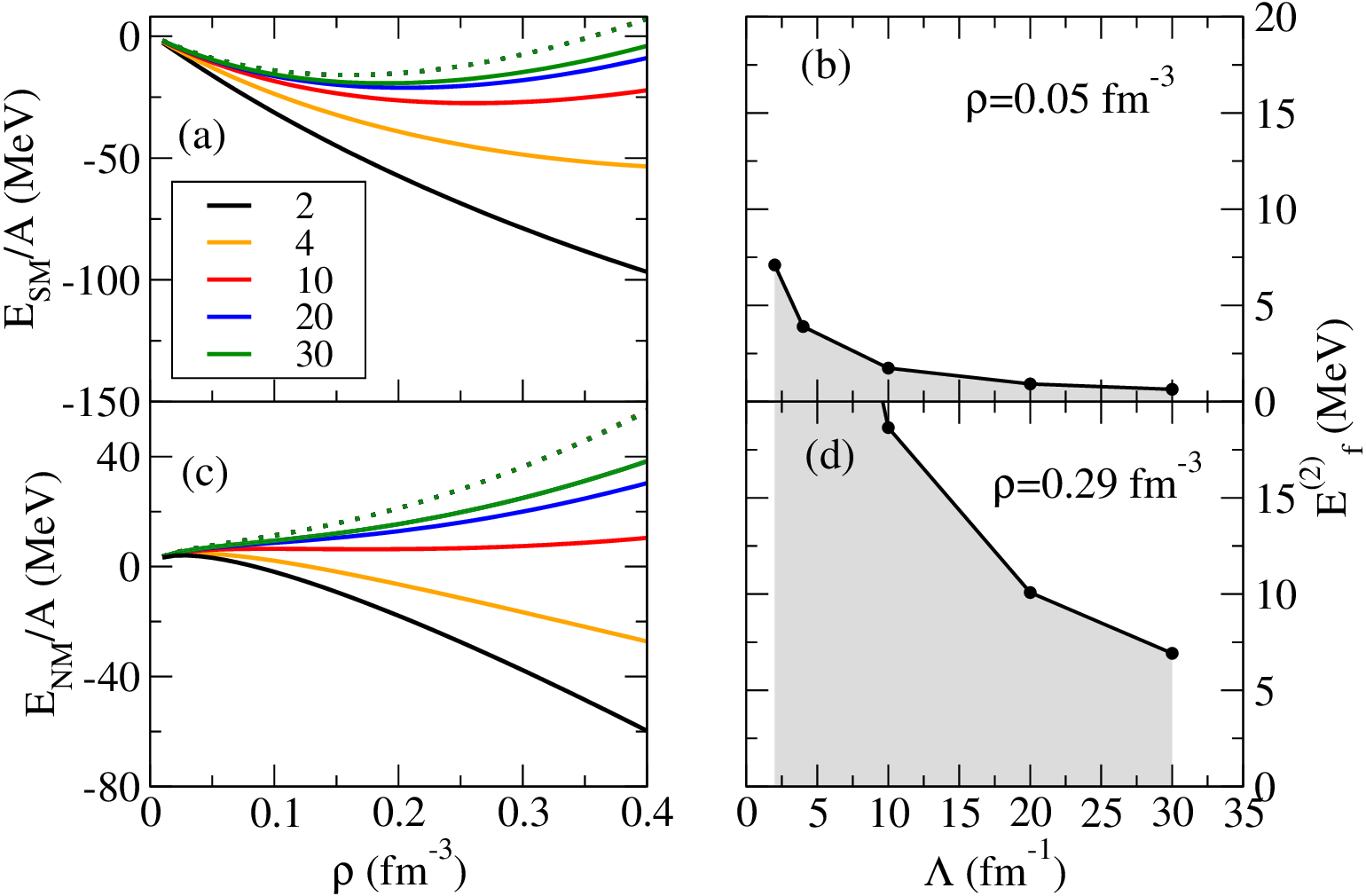}
\caption{(Color online) EOSs for SM (a) and NM (c) (dotted lines) and the same quantities computed without the finite second--order parts (solid lines). 
The cutoff values (in fm$^{-1}$) are shown in the legend. 
Panels (b) and (d) show 
the evolution of the second--order finite parts as a function of the cutoff for SM at two values of the density, 0.05 (b) and 0.29 (d) fm$^{-3}$. } 
\label{etot}
\end{figure}


Without gradient terms, our model has 5 parameters at LO ($t_0$, $t_3$, $x_0$, $x_3$, and $\alpha$) and 7 parameters at NLO, with two additional coefficients. This is enough for describing SM and NM. At this stage, we have at NLO a lower number of parameters compared to a Skyrme functional applied to matter at the mean--field level (9 parameters without the spin--orbit coupling constant). 
The Skyrme parameters are 10 when finite nuclei are treated and it is known that, with this number of parameters, Skyrme functionals are very predictive over the entire chart of nuclei. This may indicate that, to have also in our model a comparable predictive power for nuclei, it will be inevitable to have a higher number of parameters. Such an aspect can be analyzed only when applications to nuclei are done. For example, we may then realize that either we have to include gradient terms at NLO (implying then an increased number of parameters at NLO) or we have to go up to the next order NNLO (with the appearance of new parameters coming from the additional counterterms).

As mentioned above, the finite terms are suppressed as the cutoff increases. Although this is partially due to the particular choice of  
renormalization scheme, it is reassuring to observe the suppression of the sum of the second--order finite parts as presented in panels (g) and (h) 
of Fig. \ref{nmenergy}. 
This is a perturbative scheme (based on the Dyson expansion),  one prerequisite being the smallness of the expansion parameter, which is directly reflected in the finite terms. The found behavior may incidentally also explain the well--recognized global success of effective theories based on Skyrme functionals.  

Such a trend may also be seen  in Fig. \ref{etot}, panels (a) and (c) for SM and NM, respectively, where the total EOSs (dotted lines, practically superposed at all cutoff values) are compared to the EOSs computed without the second--order finite parts (solid lines) for various values of the cutoff. Incidentally, we mention that the benchmark SLy5 EOSs are also superposed to the dotted curves, as one may easily expect by looking at the values of $\chi^2$ reported in Table \ref{fitc2}.
By increasing the cutoff, the solid curves get closer to the dotted ones. As an illustration, panels (b) and (d) show the evolution of the second--order finite parts as a function of the cutoff for SM  at two values of the density, 0.05 (b) and 0.29 (d) fm$^{-3}$ which are, respectively, lower and higher than the empirical saturation density of SM. 
One may observe that the finite part contribution to the total energy is less important at $\rho=$ 0.05 fm$^{-3}$ than at 
 $\rho=$ 0.29 fm$^{-3}$. For example, the contribution at $\Lambda=$ 30 fm$^{-1}$ to the total energy is equal to 0.64 (6.92) MeV 
at $\rho=$ 0.05 (0.29) fm$^{-3}$ (the total energy is equal to -8.97 (-8.55) MeV at   $\rho=$ 0.05 (0.29) fm$^{-3}$).
If the underlying EFT--type expansion is truncated at the order $(\frac{Q}{\Lambda_{hi}})^n$, the above converged pattern is dominated by a residual cutoff--dependence of the order $(\frac{Q}{\Lambda_{hi}})^{n+1}$ \cite{harald,harald2}.
Our results show therefore that: (a) $n\ge 1$; (b) the $k_F$ corresponding to $\rho=$0.29 fm$^{-3}$ is still smaller than $\Lambda_{hi}$.

We have discussed the organizing scheme introduced in Ref. \cite{yang2017-b}  by analyzing the NLO EOSs of SM and NM. 
Such NLO EOSs are composed by renormalized first--order terms and a second--order finite part. 
We have analyzed the convergence of the renormalized parameters to cutoff--independent values and the related 
convergence of each energy contribution of the EOS. The convergence features with respect to the cutoff are found to be  different for each renormalized parameter (and, consequently, for the associated energy contribution): 
the parameters (energy contributions)
get closer to their last values (curves), corresponding to $\Lambda=$ 30 fm$^{-1}$, at cutoff values which depend on the power of $k_F$ involved in the corresponding EOS terms. 
 
 Finally, the asymptotic behavior of the parameters constrains the second--order finite parts to be progressively suppressed by increasing the cutoff, both in SM and in NM. Such a perturbative behavior has been shown to be more pronounced at lower densities. 
By analyzing the perturbativity of the proposed scheme, we have validated it as an encouraging step forward towards the definition of a power counting in EDF theories. In the absence of perturbativity we would have concluded that the proposed scheme is definitely not compatible with a possible underlying power counting.    

This opens several perspectives. The establishment of a power counting in EDF theories would be a decisive achievement, especially crucial for models where the MF approximation is overcome. Furthermore, a scheme which is even more perturbative in dilute matter  would be of particular interest when employed in specific scenarios such as exotic nuclei with diffuse surfaces or the crust of neutron stars, where densities lower than the saturation density are encountered. 

\begin{acknowledgments}
We acknowledge fruitful discussions with Denis Lacroix. This project has received funding from the European Union Horizon 2020 research and innovation program under Grant No. 654002 and from the IN2P3-CNRS BRIDGE-EDF project. 
\end{acknowledgments}

\appendix
\section{Renormalized parameters and second--order finite parts}
\label{notation}
The renormalized parameters are defined as: 
\begin{eqnarray}
t_0^{\Lambda}=t_0-\frac{m\Lambda}{2\pi^2}G,
\label{t0} 
\end{eqnarray}
\begin{eqnarray}
t_{3,\alpha}^{\Lambda}=t_{3,\alpha}-\frac{m\Lambda}{2\pi ^{2}} H,
\label{t3} 
\end{eqnarray}
\begin{eqnarray} 
t_0^{\Lambda*}=t_0^*-\frac{m\Lambda}{2\pi^2} I,
\label{capt0} 
\end{eqnarray}
and 
\begin{eqnarray}
t_{3,\alpha}^{\Lambda*}= 
t_{3,\alpha}^* - \frac{m\Lambda}{2\pi^2} L , 
\label{t3tilde}
\end{eqnarray}
where 
\begin{eqnarray}
G=t_0^2(1+x_0^2),
\label{gg} 
\end{eqnarray}
\begin{eqnarray}
H=t_{0}t_{3,\alpha}\left[ 2\left(
1+x_{0}x_{3}\right) +3 \tilde{\alpha}  \right],
\label{hh} 
\end{eqnarray}
\begin{eqnarray} 
I=(t_0^*)^2,
\label{ii} 
\end{eqnarray}
\begin{eqnarray}
L= 
2 t_0^* t_{3,\alpha}^*  \left( 1+\tilde{\alpha} \right) , 
\label{ll}
\end{eqnarray}
with $\tilde{\alpha}=\alpha(3+\alpha)/8$.
The renormalized parameters related to the counterterms are
\begin{eqnarray} 
C^{\Lambda}=V_{NLO}^{SM} -\frac{m\Lambda}{2\pi^2} P
\label{cc} 
\end{eqnarray}
and 
\begin{eqnarray} 
C^{\Lambda*}=V_{NLO}^{NM} -\frac{m\Lambda}{2\pi^2} R,
\label{cc*} 
\end{eqnarray}
where $V_{NLO}^{SM}$ and $V_{NLO}^{NM}$ are the parameters characterizing the mean--field counterterm contribution to the EOS for SM and NM, respectively, and  $P$ and $R$ are given by  
\begin{eqnarray} 
P=(t_{3,\alpha})^2 \left[ x_3^2+\left(1+\frac{3}{2}\tilde{\alpha}\right)^2\right]
\label{pp} 
\end{eqnarray}
and
\begin{eqnarray} 
R=(t_{3,\alpha}^*)^2 (1+\tilde{\alpha})^2,
\label{qq} 
\end{eqnarray}
respectively.

The expressions for the quantities $F^{(2)}$ in Eqs. (\ref{a}) and (\ref{a2}) are 

\begin{equation}
F^{(2)}_{S,1}= 3 \beta G, \; F^{(2)}_{S,2}= 3 \beta H, \;  F^{(2)}_{S,3}= 3\beta P, 
\label{ff1}
\end{equation}
\begin{equation} 
 F^{(2)}_{N,1}= \beta I, \;  F^{(2)}_{N,2}=\beta L, \;  F^{(2)}_{N,3}=\beta  R, 
\label{fff}
\end{equation}
where 
\begin{equation}
 \beta=\frac{m(11-2ln2)}{280\pi^4} 
\label{beta}
\end{equation}

\begin{center}
\begin{widetext}
\section{Adjusted parameters and $\chi^2$ values}
\label{adjust}

\begin{table}[h]
\begin{tabular}{ccccccccccccc}
\hline
\hline
$\Lambda $ (fm$^{-1}$) & 2 & 4 & 6 & 8 & 10 & 12 & 14 & 16 & 18 & 20 & 25 & 30 \\ \hline
$t_{0}$ (fm$^{2}$)  & -3.443 & -2.593 & -2.172 & -1.906 & -1.722 & -1.582 & -1.473 & -1.383 & -1.310 & -1.245 & -1.120 & -1.028 \\ 
$t_{3}$ (fm$^{2+3\alpha }$)  & 2.123 & 1.724 & 1.411 & 1.124 & 1.016 & 0.932 & 0.854 & 0.794 & 0.742 & 0.697 & 0.625 & 0.574 \\
$x_{0}$ &  0.566  & 0.564 & 0.556 & 0.554 & 0.547 & 0.542 & 0.540 & 0.537 & 0.536 & 0.534 & 0.530 & 0.527 \\
$x_{3}$ & 7.711 & 8.551 & 9.342 & 10.745 & 10.931 & 11.080 & 11.378 & 11.584 & 11.805 & 12.029 & 12.144 & 12.153 \\
$\alpha$ &  0.214 & 0.217 & 0.222 & 0.226 & 0.229 & 0.233 & 0.235 & 0.237 & 0.238 & 0.240 & 0.242 & 0.243 \\
$C^{\Lambda}$ (fm$^{2+6\alpha }$) & 0.350 & 0.329  & 0.301 & 0.287 & 0.280 & 0.277 & 0.272 & 0.266 & 0.264 & 0.262 & 0.258 & 0.263 \\
$C^{\Lambda \ast }$ (fm$^{2+6\alpha }$)  & 0.442 & 1.270 & 1.681 & 1.923 & 2.077 & 2.170 & 2.251 & 2.304 & 2.357 & 2.389 & 2.447 & 2.498 \\
$\chi ^{2}$  & 0.029  & 0.027 & 0.038  & 0.048 & 0.059  & 0.068  & 0.076 & 0.085 & 0.090  & 0.098 & 0.110  & 0.117 \\
\hline
\hline
\end{tabular}%
\caption{$\chi^2$ and adjusted parameters obtained from $\Lambda=$ 2 to 30 fm$^{-1}$.}
\label{fitc2} 
\end{table}
\end{widetext}
\end{center}

\end{document}